\documentclass[english,pra,twocolumn]{revtex4}
\usepackage{ae,aecompl}
\usepackage[T1]{fontenc}
\usepackage[latin9]{inputenc}
\usepackage{amsmath}
\usepackage{color}
\usepackage{graphicx}
\usepackage{amssymb}

\makeatletter

\usepackage{color}

\usepackage{bm}
\usepackage[normalem]{ulem}
\usepackage{comment}

\makeatletter

\def\kbar{{\mathchar'26\mkern-9mu k}}

\makeatother

\usepackage{babel}
\makeatother

\begin{document}

\title{Kicked rotor quantum resonances in position space}

\author{Maxence Lepers}

\author{V\'eronique Zehnl\'e}

\author{Jean Claude Garreau}

\affiliation{Laboratoire de Physique des Lasers, Atomes et Mol\'ecules, Universit\'e
des Sciences et Technologies de Lille, CNRS; CERLA; F-59655 Villeneuve
d'Ascq Cedex, France}

\begin{abstract}
We present an approach of the kicked rotor quantum resonances in position-space,
based on its analogy with the optical Talbot effect. This approach
leads to a very simple picture of the physical mechanism underlying
the dynamics and to analytical expressions for relevant physical quantities,
such as mean momentum or kinetic energy. The ballistic behavior, which
is closely associated to quantum resonances, is analyzed and shown
to emerge from a coherent adding of successive kicks applied to the
rotor thanks to a periodic reconstruction of the \emph{spatial} wavepacket. 
\end{abstract}
\maketitle

\section{Introduction }

The kicked rotor is a simple system that plays a central role in studies
of classical and {}``quantum chaos''. The latter is defined as the
quantum behavior of a system whose classical counterpart is chaotic.
In its simpler form, a kicked rotor (KR) is formed by a particle orbiting
a fixed circular orbit to which an instantaneous force (a \emph{kick})
is applied periodically. The corresponding classical dynamics is found
to be regular (periodic) for weak kick intensities, for intermediate
kick intensities chaotic regions develop in limited zones of the phase
space, and for strong enough forcing an ergodic diffusion appears
\citep{Chirikov:ChaosClassKR:PhysRep79}. In 1995, Raizen and co-workers
\citep{Raizen:QKRFirst:PRL95} established this system as a privileged
ground for studies of quantum chaos by realizing experimentally a
kicked rotor in the quantum regime with laser-cooled atoms placed
in a kicked (i.e. rapidly turned on and off) laser standing wave.

Despite its apparent simplicity, the quantum kicked rotor (QKR) has
very remarkable dynamical properties. One of the most studied is the
so-called {}``dynamical localization'': in contrast to the classical
case, the ergodic diffusion does not last forever in the quantum case.
After a characteristic {}``localization time'', the diffusion is
stopped by destructive quantum interferences \citep{Casati:LocDyn:79,Izrailev:LocDyn:PREP90,Raizen:QKRFirst:PRL95}.
Another feature of the QKR that has been the object of a recent burst
of theoretical and experimental activity is the existence of \emph{quantum
resonances} (QRs), whose most dramatic manifestation is the appearance,
for specific values of the parameters, of a \emph{ballistic} motion,
instead of a diffusive or a localized behavior. Quantum resonances
are the main subject of the present paper. 

The experimental realization of the QKR has triggered in the last
decade an impressive number of studies involving dynamical localization
\citep{Raizen:LDynNoise:PRL98,Christ:LDynNoise:PRL98,AP:Bicolor:PRL00,
AP:SubFourier:PRL02,Monteiro:LocDelocDoubleKick:PRL06,AP:Reversibility:PRL05,AP:PetitPic:PRL06},
quantum transport \citep{Monteiro:DoubleKick:PRL04,Monteiro:DirectedMotionDKR:PRL07,
Leonhardt:KREarlyTimeDiff:PRL04,Leonhardt:TransportKR:PRL05},
ratchets \citep{Monteiro:RatchetOptLatTh:PRL02,Renzoni:RatchetOptLat:PRL04,Arimondo:RatchetOptLat:PRA06},
chaos-assisted tunneling \citep{Raizen:ChaosAssistTunnel:Science01,Rubins-Dunlop:ChaosAssitTunnel:Nature01},
classical and quantum resonances \citep{Raizen:KRClassRes:PRL98,Philips:QuantumResTalbot:PRL99,
DArcy:AccModes:PRL99,Fishman:KRQuantumRes:PRL02}.
Quantum resonances have been used in studies of fundamental aspects
of quantum chaos such as quantum stabilization \citep{DArcy:QuantumStability:PRL03}
or measurements of the gravitation \citep{DArcy:GravityQuantRes:PRL04}.
{}``High-order'' quantum resonances were also observed recently
\citep{DArcy:HighOrderQRes:PRL03,Phillips:HighOrderQuantResBEC:PRL06,Steinberg:HighOrderQRes:PRL07}
both with a Bose-Einstein condensate and laser-cooled atoms.

In the present work, we show that the kicked rotor quantum resonances
have a very simple and intuitive interpretation in \emph{position}
space, as opposed to the more common momentum space representation
(see \citep{Wimberger:QuantRes:NL06} and references therein). QRs
in position space have been previously considered by Izrailev and
Shepelyansky \citep{Shepel:QuantResPosSpace:TMP80,Izrailev:LocDyn:PREP90}.
Here, we extend their approach and build a simple physical picture
that enlightens the underlying physics and allows the calculation
of quantities of experimental relevance, such as average momentum
or the kinetic energy.

\section{Physical origin of quantum resonances\label{sec:Physical-origin}}

In this section we describe the fundamental properties of the kicked
rotor dynamics and discuss the physical origin of quantum resonances.
We aim at a simple presentation that puts into evidence the physical
mechanisms even for the reader unfamiliar with quantum resonances.
Some aspects of the following discussion may be considered as trivial,
but the discussion is nevertheless necessary to introduce ideas and
notation non-ambiguously.

The atom-optics realization of a (quantum) kicked rotor consists in
placing laser-cooled atoms in a far-detuned laser standing wave. In
such conditions, the atoms feel the light intensity as a mechanical
potential affecting their center-of mass degree of freedom \citep{CCT:Houches:90,
MetcalfStratenLaserCooling,MeystreAtOpt}.
The standing wave is pulsed periodically, being on for a time interval
so short that the motion of the atoms can be neglected; in such conditions,
the pulses can be considered as delta functions (kicks). The standing
wave sinusoidal modulation of intensity generates a spatial potential
in $\sin\left(2k_{L}x\right)$, where $k_{L}=2\pi/\lambda_{L}$ is
the wavenumber of the radiation forming the standing wave. As the
de Broglie wavelength of laser-cooled atoms is about $\lambda_{L}/3$,
it is comparable to the periodicity of the potential, $\lambda_{L}/2$,
and the system is in the quantum regime, provided that decoherence
is negligible during the experiment. In practice, this implies using
a far-detuned radiation to reduce spontaneous emission to acceptable
levels.

We shall use a normalized spatial coordinate $X=2k_{L}x=2x/\lambda_{L}$
which plays in fact the role of a cyclic variable: the spatial periodicity
of the potential implies that the physics is the same if translations
by a multiple of $\lambda_{L}/2$ are performed. We can thus use either
a {}``linear'' (``unfolded'') representation of the KR, using the
linear variable $X$ or a {}``cyclic'' or {}``folded'' representation
using the angular variable $\theta=X[\mathrm{mod}\;2\pi]$ ($x[\mathrm{mod}\;\lambda_{L}/2]$
in usual units) \citep{note:localized}.

Choosing, moreover, units such that the mass of the particle is 1
and the time-period of the forcing is $T=1$, the Hamiltonian of the
KR reads\begin{equation}
H=\frac{P^{2}}{2}+K\cos X\sum_{n=0}^{N-1}\delta(t-n)\label{eq:HKR}\end{equation}
where $P$ is the momentum (scaled by a factor $M/2k_{L}T$ ) $X$
the position in the periodic potential, $K$ (usually called {}``stochasticity
parameter'') the intensity of the kicks, and $n$ a discrete time
corresponding to the $n^{th}$ kick ($n$ is an integer in normalized
units).

Labeling $X_{t},P_{t}$ the position and momentum immediately \emph{after}
the $t^{th}$ kick, and integrating the classical Hamilton equations
of motion corresponding to Eq.~(\ref{eq:HKR}) produces the so-called
Chirikov's {}``standard map'' \citep{Chirikov:ChaosClassKR:PhysRep79}:\begin{eqnarray}
X_{t+1} & = & X_{t}+P_{t}\nonumber \\
P_{t+1} & = & P_{t}+K\sin X_{t+1}.\label{eq:MapChirikov}\end{eqnarray}
The classical dynamics is found to be periodic below a critical value
of the kick intensity $K_{c}\approx0.9716$. Chaotic regions appear
in phase-space for $K>K_{c}$, and progressively grow as $K$ increases.
For $K\gtrsim5$ the islands of stable dynamics are barely visible,
and the classical dynamics becomes an ergodic diffusion in phase space.
The average kinetic energy then increases linearly with time (or kick
number $t$): $\left\langle P^{2}\right\rangle /2=D_{c}t$, where
$D_{c}\approx K^{2}/2$ is a diffusion coefficient that can be explicitly
calculated \citep{Rechester:KRDiffCoeff:PRA81}.

The standard map presents well-known \emph{classical} resonances,
also called \emph{accelerator modes}. For example, set $K=2\pi$,
and consider a particle with initial conditions $X_{0}=\pi/2$ and
$P_{0}=0$. Iteration of Eqs. (\ref{eq:MapChirikov}) shows that $P_{t}=2\pi t$
and $X_{t}=2\pi(t^{2}-t)+\pi/2$. The momentum increases linearly
with time and the kinetic energy $P^{2}/2$ increases \emph{quadratically}
with time, which is a signature of a \emph{ballistic} dynamics, in
contrast with the linear increase in the diffusive case. The origin
of the ballisticity is easily seen: for this particular choice of
$K$ and of the initial conditions, the particle is always kicked
at the \emph{same} position (modulus $2\pi$), that is $\sin X_{t}$$=\sin$$\left(2\pi(t^{2}-t)+\pi/2\right)$$=1$,
and thus receives the same amount of linear momentum per kick. For
arbitrary values of $K$ or of the initial conditions, the particle
is kicked in different positions, and the effect of some kicks compensate
the effect of other kicks, leading to a \emph{slower} increase of
the energy. One can easily convince oneself that, in general, there
is a classical resonance for $K=2\pi p$ with integer $p$ and adequate
initial conditions. The classical resonance was experimentally observed
in the atom-optics realization of the KR \citep{Raizen:KRClassRes:PRL98}.
There are also classical \emph{antiresonances}: e.g. for the above
initial condition and $K=3\pi/2$, successive kicks have \emph{opposite}
directions, and the momentum jumps endlessly between two values, $P_{t}=0$
and $P_{t+1}=3\pi/2$.

In order to study the quantum dynamics of the KR, one considers the
one-period evolution operator, also called Floquet operator:
\begin{equation}
U=e^{-iH/\kbar}=\exp\left(-i\frac{K}{\kbar}\cos X\right)\exp
\left(-i\frac{P^{2}}{2\kbar}\right)\label{eq:FloquetOperator}
\end{equation}
where $\kbar=4\hbar k_{L}^{2}T/M$ is the normalized Planck's constant
resulting from the definition of normalized variables satisfying the
commutation relation $[X,P]=i\kbar$; it thus describes the {}``quanticity''
of the system ($\kbar$$\rightarrow0$ is the classical limit). The
above operator relates the quantum state after the $t^{th}$ kick
to the quantum state after the $(t-1)^{th}$ kick \citep{note:conventions}.
It is a remarkable fact that this operator factorizes into the product
of two exponentials, a free evolution followed by the kick effect,
despite the fact that $X$ and $P$ do not commute. This is a consequence
of the $\delta$-function time dependence: during the instantaneous
kick, the evolution related to the kinetic energy term is negligible.
If one starts e.g. in the $P$-representation, the free evolution
(from $t^{+}$ to $(t+1)^{-}$) corresponds to simply adding a phase.
One can then convert to the $X$-representation where the kick operator
is diagonal and also simply adds a phase; one then goes back to the
$P$-representation. The calculation of the quantum evolution over
a period thus {}``costs'' only two Fourier transforms and two multiplications.
It is this formal and numerical simplicity that made the QKR so popular.
Despite this simplicity, the cross-action of these two operators makes
the QKR dynamics very rich: the free evolution operator mixes space
components and the kick operator adds new momentum components, generating
a complex quantum-interference pattern.

An important property of the kick operator is that it is periodic
in space. This means that its eigenstates have a Bloch-wave (BW) structure.
This is formally seen by developing it in terms of Bessel functions
(noting that $e^{inX}$ is the momentum translation operator $e^{inX}\left\vert P\right\rangle
 =\left\vert P+n\kbar\right\rangle $)
\[
\exp\left(-i\kappa\cos X\right)=\sum_{m=-\infty}^{+\infty}\left(-i\right)^{m}J_{m}(\kappa)e^{imX}\]
where we introduced the quantity
\begin{equation}
\kappa\equiv\frac{K}{\kbar}.\label{eq:Kappa}
\end{equation}
Therefore, if one starts from an initial state of well-defined momentum
$\left\vert P_{0}\right\rangle $, only states of the form $\left\vert P_{0}+m\kbar\right\rangle $
($m$ \emph{integer}) will appear in the dynamics. We can write any
arbitrary momentum $P_{0}$ in the form\[
P_{0}=(m+\beta)\kbar\]
with $m$ integer and $\beta\in[-1/2,1/2)$ (i.e. $\beta$ is in the
{}``first Brillouin zone''). This introduces the \emph{quasimomentum}
$\kbar\beta$ which, as just demonstrated, is a constant of motion.
The particle wavefunction $\psi(X)=\left\langle X\right|\left.\psi\right\rangle $
can be written as
\[
\psi(X)=\frac{1}{\sqrt{2\pi}}\int_{-\infty}^{\infty}e^{ikx}\widetilde{\psi}(k)dk,\]
where $k=P/\kbar$, and can thus be decomposed in:\begin{eqnarray}
\psi(X) & = & \frac{1}{\sqrt{2\pi}}\int_{-1/2}^{+1/2}d\beta\sum_{m=-\infty}^{+\infty}\widetilde{\psi}_{\beta}(m)e^{i(m+\beta)X}\nonumber \\
 & = & \int_{-1/2}^{+1/2}d\beta\psi_{\beta}(X),\label{eq:decompo-beta}\end{eqnarray}
where we introduced the quasimomentum component $\psi_{\beta}(X)$:\begin{equation}
\psi_{\beta}(X)=\frac{1}{\sqrt{2\pi}}\sum_{m=-\infty}^{^{\infty}}\widetilde{\psi}_{\beta}(n)e^{i(m+\beta)X}\label{eq:PsiBetaXFunction}\end{equation}
with\begin{equation}
\psi_{\beta}(X)=e^{i\beta X}u_{\beta}(X).\label{eq:ubeta}\end{equation}
 The function \begin{equation}
u_{\beta}(X)=\frac{1}{\sqrt{2\pi}}\sum_{m=-\infty}^{^{\infty}}\widetilde{\psi}_{\beta}(m)e^{imX}\label{eq:ubeta_bis}\end{equation}
is of a $2\pi$-periodic function. Eq.~(\ref{eq:PsiBetaXFunction})
is a direct manifestation of the Bloch theorem: the particle is described
by a Bloch wave, $\psi_{\beta}(X)$, which is the product of a plane-wave
of well-defined quasimomentum $\beta$ and a periodic function $u_{\beta}(X)$.

In most experimental realizations of the kicked rotor with laser-coled
atoms the initial velocity distribution is larger than the Brillouin
zone of the system and all quasimomenta are present; it may thus be
necessary to average observable quantities over the quasimomentum.
It is shown in the App. \ref{app:mean} that these averages values
can be simply expressed in terms of the BW decomposition:\begin{subequations}
\label{eq:EP-moy-}\begin{eqnarray}
\langle P\rangle(t) & = & \int_{-1/2}^{1/2}d\beta\langle P\rangle_{\beta}\label{eq:moyenne-P-psi}\\
\langle E\rangle(t) & = & \int_{-1/2}^{1/2}d\beta\langle E\rangle_{\beta}\label{eq:moyenne-E-psi}\end{eqnarray}
 \end{subequations}where the subscript $\beta$ indicates an average
over $\psi_{\beta}$.

\begin{figure}
\begin{centering}
\includegraphics[bb=50bp 20bp 410bp 302bp,clip,width=8cm]{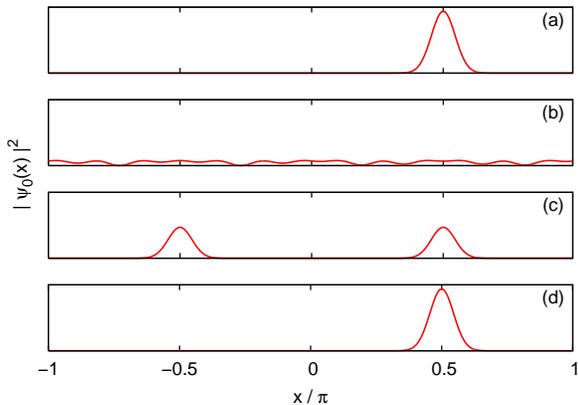} 
\par\end{centering}

\caption{\label{fig:FreeEvolution} Free evolution of a localized wavepacket
($\beta=0)$between two successive kicks for $\kbar=4\pi$. (a) Initial
wavepacket centered at $X=\pi/2$ shown in the interval $X\in\left[-\pi,\pi\right]$,
(b) wavepacket at $t=0.115$, showing complete delocalization, (c)
wavepacket at $t=0.25$, showing two replicas of the initial shape,
(d) {}``reconstructed'' wavepacket at $t=1^{-}$, identical to the
initial one. }

\end{figure}

In order to unveil the physical origin of quantum resonances, let
us take, for simplicity, an initial state of quasimomentum $\beta=0$
and apply to it the free evolution $\exp(-i\kbar P^{2}/2)$:\begin{widetext}\begin{eqnarray}
\psi_{0}(X,t=1^{-}) & = & \frac{1}{\sqrt{2\pi}}\sum_{m}\widetilde{\psi}_{0}\left(m,t=0\right)\exp\left(-i\kbar\frac{m^{2}}{2}\right)\exp\left(imX\right).\label{eq:SF-theta}\end{eqnarray}
\end{widetext}If we set in the above expression \[
\frac{\kbar}{2}=2\pi\ell,\]
with $\ell$ arbitrary positive integer, we see that the argument
of the exponential phase factor in Eq.~(\ref{eq:SF-theta}) is always
an integer multiple of $2\pi$. This means that \emph{the free evolution
over a period leaves the wavepacket invariant}. Fig.~\ref{fig:FreeEvolution}
shows an initial BW that is spatially localized \citep{note:combs}
in the first Brillouin zone and which evolves freely with time. The
BW first becomes completely delocalized but, as seen above, it {}``focalizes''
back to its initial shape just before the kick! Our using of an optics
language is more than fortuitous: a monochromatic beam diffracted
by a grating also reforms after a certain propagation length, an effect
known in optics as the {}``Talbot effect''. The quantum resonance
is the atom-optics analog of the optical Talbot effect \citep{Philips:QuantumResTalbot:PRL99}.

The only possible evolution comes from the kicks, whose effect is
to add new momentum components ({}``sidebands'') separated by a
multiple of $\kbar$. As the effect of the kick is diagonal in position
representation; it adds a position-dependent phase to the wavefunction:\[
\psi_{0}(X,t=1)=\exp\left(-i\kappa\cos X\right)\psi_{0}(X,t=0).\]
As the wavepacket has always the same shape when the kick is applied,
it acquires the \emph{same} phase from each kick, and the kick effects
add \emph{coherently}. The quantum resonance condition is thus a constructive
interference effect (in contrast with dynamical localization for instance,
which results from destructive interferences that {}``freezes''
the wavefunction evolution).

The above discussion puts into evidence an analogy (despite its very
different nature) between quantum and classical resonances: in both
cases, for particular values of the parameters, the dynamics is such
that the particle (the wavepacket in the quantum case) is always kicked
at the same position and the effect of the kicks adds (constructively
interferes) to produce a linear increase of the momentum. An important
difference, however, is that in the classical case the effect is related
to the intensity of the kicks -- the resonance condition thus depends
on the parameter $K$ -- whereas in the quantum case it is related
to the constructive accumulation of quantum phases and the resonance
condition thus depends on the value of $\kbar$.

In the following sections we shall develop an approach to calculate
and interpret the behaviors of the system for different types of QRs.
In section \ref{sec:SimpleQuantumResonances}, the so-called {}``simple''
resonances'' corresponding to $\kbar=2\pi\ell$ ($\ell$ positive
integer) are analyzed, in which the initial wavepacket {}``refocuses''
in a single replica of the initial wavepacket. In section \ref{sec:PiQuantumRes},
we focus on the so-called {}``high-order resonances'' ($\kbar=4\pi r/s$,
$r,s$ integers), where the free evolution of an initial wavepacket
generates various spatially-separated replicas of the initial wavepacket.

\section{Analysis of ``simple'' quantum resonances\label{sec:SimpleQuantumResonances}}

Quantum resonances obeying the condition $\kbar=2\pi\ell$ (with $\ell$
a positive integer) are named {}``simple'' quantum resonances (SQR).
With this form for $\kbar$, it is shown that the shape of BW remains
invariant in the free propagation step between two kicks and that
the effects of the kicks may lead to a ballistic (linear) growth of
the momentum; the kicking period is a multiple of the half-Talbot
time defined in classical optics, which is the condition for the optical
(integer) Talbot effect \citep{Berry:FracTalbotEffect:JMO96,Berry:TalbotEffect:JMO99}.

\subsection{Time-evolution\label{sub:Time-evolution} }

Let us consider, at some (integer) time $(t-1)$ a state $\psi_{\beta}(X,t-1)$
corresponding to the general form of Eq.~(\ref{eq:PsiBetaXFunction}),
and apply to it the free-evolution operator with $\kbar=2\pi\ell$
:\begin{widetext}\begin{eqnarray*}
\exp\left(-i\frac{P^{2}}{2\kbar}\right)\psi_{\beta}(X,t-1) & = & 
\frac{1}{\sqrt{2\pi}}\sum_{m}\widetilde{\psi}_{\beta}(m,t-1)
\exp\left(-i\pi\ell\left(m+\beta\right)^{2}\right)e^{i(m+\beta)X}\\
 & = & \frac{1}{\sqrt{2\pi}}\exp\left(-i\pi\ell\beta^{2}\right)\sum_{m}
\widetilde{\psi}_{\beta}(m,t-1)\exp\left(-i\pi\ell m^{2}\right)
\exp\left(-i2\pi\ell m\beta\right)e^{i(m+\beta)X}\end{eqnarray*}
As $\ell m^{2}$ and $\ell m$ have the same parity, the first exponential
factor under the sum is equal to $\exp\left(-i\pi\ell m\right)$ ;
it can thus be combined with the second term, yielding $\exp\left(-im\kbar\left(\beta+1/2\right)\right)$.
We obtain\begin{eqnarray*}
\psi_{\beta}(X,t^{-}) & = & \frac{1}{\sqrt{2\pi}}\exp\left(-i
\frac{\kbar\beta^{2}}{2}\right)\sum_{m}\widetilde{\psi}_{\beta}(m,t-1)
\exp\left(-im\kbar\beta^{\prime}\right)e^{i(m+\beta)X}\end{eqnarray*}
\end{widetext}with $\beta^{\prime}\equiv\beta+1/2$. This last expression
can be rewritten, using Eq.~(\ref{eq:PsiBetaXFunction}), as \begin{equation}
\psi_{\beta}\left(X,t^{-}\right)=\exp\left(i\frac{\kbar\beta(\beta+1)}{2}
\right)\psi_{\beta}\left(X-\kbar\beta^{\prime},t-1\right).\label{eq:psi-resonance}\end{equation}
Applying now the kick operator (which is diagonal in the $X$ representation),
we obtain a recurrence relation linking the wavepackets at times $t$
and $(t-1)$:\begin{equation}
\psi_{\beta}(X,t)=e^{-i\kappa\cos X}\exp\left(i\frac{\kbar\beta(\beta+1)}{2}\right)
\psi_{\beta}\left(X-\kbar\beta^{\prime},t-1\right).\label{eq:2pil-rec-psi}\end{equation}
The above result shows that in the conditions of a simple resonance
$\left|\psi(X,t)\right|^{2}=\left|\psi(X-v,t-1)\right|^{2}=\left|\psi(X-vt,0)\right|^{2}$:
the square-modulus of the wavefunction remains invariant immediately
before the kick, except for a drift with a {}``velocity'' \citep{note:drift}
\begin{equation}
v=\kbar\left(\beta+\frac{1}{2}\right).\label{eq:v}\end{equation}
Moreover, the wavefunction acquires a position-dependent phase due
to the kick.

By iterating Eq.~(\ref{eq:2pil-rec-psi}) down to $t=0$, we can express
$\psi_{\beta}(x,t)$ in terms of the initial wavefunction\begin{equation}
\psi_{\beta}(X,t)=\exp\left(-i\kappa\Phi(X,t)\right)\psi_{\beta}\left(X-vt,0\right),\label{eq:psi-beta-evolution}\end{equation}
where the accumulated phase is given by \citep{note:factor} \begin{equation}
\Phi(X,t)=\sum_{s=0}^{t-1}\cos\left(X-vs\right).\label{eq:phase}\end{equation}

\begin{figure}
\begin{centering}
\includegraphics[bb=50bp 20bp 410bp 302bp,clip,width=8cm]{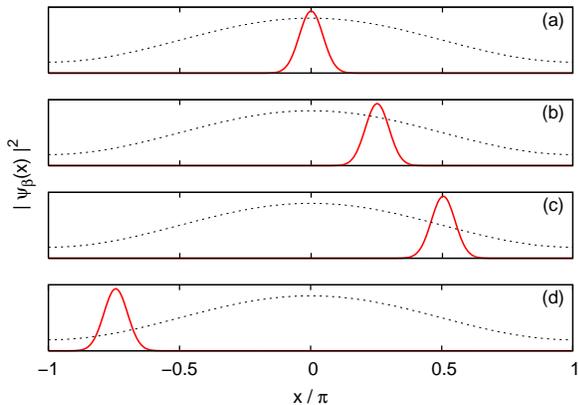} 
\par\end{centering}

\caption{\label{fig:qi-irrat}Wavepacket plotted at different integer times,
for $\kbar=4\pi$ and an \emph{irrational} quasimomentum ($\beta=\frac{\pi}{50}\approx0.063$).
(a) Initial wavepacket ($|\psi_{\beta}|^{2})$ centered at $X=0$,
(b) $t=1$, (c) $t=2$, (d) $t=5$. The shape of the potential is
represented in dotted lines. The wavepacket interacts with the potential
at different positions, and the average effect tends to zero.}

\end{figure}

A remarkable property is obtained if \begin{equation}
v=2\pi\frac{p}{q}\label{eq:reso_antireso}\end{equation}
with $p,$$q$ integer, e.g. for rational values of the quasimomentum.
Then, the phase defined in Eq.~(\ref{eq:phase}) takes the values
$\Phi(X,t=q)=0$ if $q\neq1$ and $\Phi(X,t=q)=\cos X$ if $q=1$.
Let us first consider the case $q\neq1$. After a recurrence time
$t_{r}=q$ the BW is given by:\[
\psi_{\beta}(X,t_{r})=\psi_{\beta}\left(X-2\pi p,0\right)=e^{i2\pi p\beta}\psi_{\beta}(X,0),\]
and the particle comes back to its initial state after $t_{r}$ kicks,
leading to a \emph{periodic} evolution. To show the physical origin
of this periodicity, let us take the simple case $v=\pi$, corresponding
to $t_{r}=2$. From Eqs. (\ref{eq:psi-beta-evolution}) and (\ref{eq:phase}),
the BW evolution over two successive kicks is \[
\psi_{\beta}(X,t=2)=e^{-i\kappa\cos X}\psi_{\beta}\left(X-\pi,t=1\right)\]
and\[
\psi_{\beta}(X-\pi,t=1)=e^{i\kappa\cos X}\psi_{\beta}(X-2\pi,t=0),\]
which means that the phase added by the kicks simply cancels after
two kicks. An analogous cancellation happens for any other value of
the velocity $v$ obeying Eq.~(\ref{eq:reso_antireso}) with $q\neq1$:
the force acting on the particle averages to zero after $t_{r}$ kicks,
and the motion of the wavepacket is a simple oscillation of period
$t_{r}$. This behavior, that cannot lead to a ballistic behavior,
is called {}``anti-resonance''.

If $v=2\pi p$ {[}or $q=1$ in Eq.~(\ref{eq:reso_antireso})], corresponding
to \[
\beta=\left(\frac{p}{\ell}-\frac{1}{2}\right)\]
the evolution of the BW is given by\begin{eqnarray*}
\psi_{\beta}(X,t_{r}=1) & = & e^{-i\kappa\cos X}\psi_{\beta}\left(X-2\pi p,0\right)\\
 & = & e^{-i2\pi p\beta}e^{-i\kappa\cos X}\psi_{\beta}(X,0).\end{eqnarray*}
The wavepacket exactly recovers its initial shape after each kick
and the particle is subjected to an \emph{identical} potential at
\emph{each} kick. In contrast with to the former case, the kicks will
add in a \emph{coherent} way, which, as we shall demonstrate in Sec.
\ref{sub:BlochWaveMeaValues}, causes a linear increase of the average
momentum, or ballistic behavior. This is the quantum resonance.

If Eq.~(\ref{eq:reso_antireso}) is not fulfilled, our picture still
allows to guess the general shape of the asymptotic evolution for
large numbers of kicks ($t\gg1).$ If $v/2\pi$ is not a rational
number, the phase $\Phi(X,t)$ will tend to 0 as $t\rightarrow+\infty$.
Once again, the contribution of the successive kicks will average
to zero and the wavepacket just {}``drifts'' (for stroboscopic times)
while keeping its initial shape (see Fig.~\ref{fig:qi-irrat}).

\subsection{\label{sub:BlochWaveMeaValues}Averages}

Let us now focus on the time evolution of average values. From Eq.
(\ref{eq:2pil-rec-psi}) we can easily obtain a recurrence relation
for the BW average position \begin{equation}
\left\langle X\right\rangle (t)=\left\langle X\right\rangle (t-1)+v\;.\label{eq:theta-recurrence}\end{equation}
A recurrence relation for the average momentum is obtained using Eq.
(\ref{eq:2pil-rec-psi}):\[
\left\langle P\right\rangle _{\beta}(t)=\left\langle P
\right\rangle _{\beta}(t-1)+K\int_{-\pi}^{\pi}dX\sin X\left|\psi_{\beta}(X-v,t-1)\right|^{2}\]
\begin{equation}
=\left\langle P\right\rangle _{\beta}(t-1)+K\int_{-\pi}^{\pi}dX\sin(X+vt)
\left|\psi_{\beta}(X,t=0)\right|^{2}.\label{eq:2pil:recurenceP}\end{equation}
This expression can be iterated down to $t=0$, leading to:\begin{widetext}\begin{eqnarray}
\left\langle P\right\rangle _{\beta}(t) & = & \left\langle P\right
\rangle _{\beta}(0)+K\left(\sum_{n=1}^{t}\int_{-\pi}^{\pi}dX\sin(X+nv)
\left|\psi_{\beta}(X,0)\right|^{2}\right)\nonumber \\
 & = & \left\langle P\right\rangle _{\beta}(0)+K\frac{\sin(tv/2)}{\sin(v/2)}
\textrm{Im}\left(e^{i(t+1)v/2}\int_{-\pi}^{\pi}dXe^{iX}\left|\psi_{\beta}(X,0)\right|^{2}\right).
\label{eq:2pil-rec-P}\end{eqnarray}
\end{widetext}The role of interferences is clearly seen in this last
result: indeed, if $v=\kbar(\beta+\frac{1}{2})=2\pi$ {[}mod $2\pi$],
the momentum increases linearly with the (stroboscopic) time $t$,
i.e\begin{equation}
\left\langle P\right\rangle _{\beta}(t)=\left\langle P\right\rangle _{\beta}(0)+Dt\label{eq:P_diffusion}\end{equation}
where the slope\begin{equation}
D=K\int_{-\pi}^{\pi}dx\sin x\left|\psi_{\beta}(x,0)\right|^{2}\label{eq:2pil-diffusion}\end{equation}
clearly appears as a force averaged over the initial spatial distribution.
The observed momentum transport shows that well-controlled diffusion
can be obtained by a suitable choice of initial conditions {[}i.e
of $\psi_{\beta}(x,0)$].

For $v=\kbar(\beta+\frac{1}{2})=\pi$ {[}mod $2\pi$], destructive
interference occurs. This can be seen from Eq.~(\ref{eq:2pil-rec-P}):
$\left\langle P\right\rangle _{\beta}(t)=\left\langle P\right\rangle _{\beta}(0)$
($t$ even) or $\left\langle P\right\rangle _{\beta}(t)=\left\langle P\right\rangle _{\beta}(0)-D$
($t$ odd). In the general case ($v\neq2\pi${[}mod $2\pi]$), the
momentum is frozen around its initial value. 

Quantum resonance effects can also be analyzed through the temporal
evolutions of the average kinetic energy. Using Eq.~(\ref{eq:psi-beta-evolution}),
one obtains:\begin{widetext}\begin{equation}
\langle E\rangle_{\beta}(t)=\langle E\rangle_{\beta}(t=0)+\frac{K^{2}}{2}
\int_{-\pi}^{\pi}dX\left(\sum_{n=1}^{t}\sin(X+nv)\right)^{2}\left|
\psi_{\beta}(X,t=0)\right|^{2}+K\int_{-\pi}^{\pi}dX\left(\sum_{n=1}^{t}
\sin(X+nv)\right)J(X,t=0)\label{eq:2pil-rec-ener}\end{equation}
where we introduced the current\[
J(X,t)=i\frac{\kbar}{2}\left(\psi_{\beta}(X,t)\partial_{X}\psi_{\beta}^{*}(X,t)-c.c.\right).\]
The constructive interference case $v=2\pi$ {[}mod $2\pi$] then
leads to an average kinetic energy increasing quadratically with time:\begin{eqnarray}
\left\langle E\right\rangle _{\beta}(t) & = & \left\langle E\right\rangle _{\beta}(t=0)
+\frac{1}{2}K^{2}t^{2}\int_{-\pi}^{\pi}dX\sin^{2}X\left|\psi_{\beta}(X,t=0)\right|^{2}
+Kt\int_{-\pi}^{\pi}dX\sin XJ(Xx,t=0)\label{eq:2pil-resonance}\end{eqnarray}
\end{widetext}which is the quantum-mechanical analog of the ballistic
motion observed -- in different conditions -- for a classical resonance.
The ballistic growth is seen to be proportional to the quantum average
of the square of the force $K^{2}\sin^{2}X$.

\subsection{Map}

Inspection of Eqs. (\ref{eq:theta-recurrence}) and Eq.~(\ref{eq:2pil:recurenceP})
suggests that the dynamics of position and momentum averages of a
Bloch-wave can be described by a map. In the following, we assume
that the initial BW is sharply localized around its mean position
$\left\langle X\right\rangle _{\beta}(t=0)=X_{0}$. Noting $P_{t}=\left\langle P\right\rangle _{\beta}(t)$
and $X_{t}=\left\langle X\right\rangle _{\beta}(t)$ we have from
Eq.~(\ref{eq:2pil:recurenceP}), $P_{t}=P_{t-1}+K\sin(X_{0}+vt)$
(with the normalization condition $\int_{-\pi}^{\pi}dX\left|\psi_{\beta}(X,t=0)\right|^{2}=1)$.
We then obtain

\begin{subequations} \label{eq:2pil-MapClassiq} \begin{eqnarray}
P_{t} & = & P_{t-1}+K\sin X_{t}\label{eq:2pil-MapClassiq-P}\\
X_{t} & = & X_{t-1}+v,\label{eq:2pil-MapClassiq-theta}\end{eqnarray}
\end{subequations}which evokes the classical map (see Eq. \ref{eq:MapChirikov}),
with the important difference that the position in Eq.~(\ref{eq:2pil-MapClassiq-theta})
does not depend on $P_{t}$ but solely on the drift velocity $v$.
This produces after $t$ kicks\begin{subequations}\begin{eqnarray*}
X_{t} & = & X_{0}+vt\\
P_{t} & = & P_{t-1}+K\sin(X_{0}+vt).\end{eqnarray*}
 \end{subequations}Iterating in turn this momentum equation produces\[
P_{t}=P_{0}+K{\textstyle \sum_{n=1}^{t}}\sin(X_{0}+nv).\]

Ballisticity is found if $v=2\pi$ (mod $2\pi$): \[
P_{t}=P_{0}+tK\sin X_{0}=P_{0}+Dt\]
with \begin{equation}
D=K\sin X_{0}.\label{eq:2pil-D}\end{equation}
For a localized packet, ballisticity emerges, as in the classical
case, if the particle is always kicked at the same position. In the
classical case, this is possible if $K=2\pi p$ ($p$ integer) and
$X_{0}=\pm\pi/2$. In the quantum case, the resonance condition depends
on the quantum parameters $\kbar$ and $\beta$ (through $v$), but
the kick intensity $K$ and the initial position $X_{0}$ determine
only the growth rate of the average momentum.

Recalling that $t$ is an integer (stroboscopic) variable counting
the kicks, one can find also a whole class of periodic behaviors.
An example is $\kbar=2\pi$ and $\beta=0$ ($v=\pi)$, with a recursion
time $t_{r}=2$, which has already been analyzed in sect. \ref{sub:Time-evolution}.
More generally, for rational quasimomentum values $\beta=p/q$ ($p,q$
integer), after a number of kicks $t_{r}$, the average momentum and
kinetic energy come back to their initial values. In such cases the
dynamics is periodic, due to the effect of kick compensation discussed
above.

\section{The $\kbar=\pi$ high-order resonance \label{sec:PiQuantumRes}}

{}``High-order'' QRs are the quantum-mechanical analogs of the \emph{fractional}
optical Talbot effect \citep{Berry:FracTalbotEffect:JMO96}. In contrast
with the situation found in simple resonances, after a free propagation
the initial packet does not reconstruct in an identical packet, but
forms two or more replicas of the original one. The action of the
kick in these subpackets generates different quantum phases and produces
quantum interference effects during the subsequent free propagation.
This makes high-order quantum resonances fundamentally different from,
and more complex than, simple ones.

High-order quantum resonances (HQRs) correspond to a dimensionless
Planck's constant of the form $\kbar=4\pi r/s$, with $r$ and $s>2$
integers. A calculation similar to that leading to Eq.~(\ref{eq:2pil-rec-psi})
shows that after the free propagation the initial packet refocalizes
into $s$ uniformly spaced subpackets if $s$ is odd, and $s/2$ subpackets
if $s$ is even. We shall consider here only the the simplest case
$\kbar=\pi$. The method presented below can in principle be generalized
to more complicated cases, but the algebra involved quickly gets very
cumbersome. As in the preceding section, our approach gives a simple
picture of the physical mechanism of HQRs.

Using the general expression for BW at time $(t-1)$, Eq.~(\ref{eq:PsiBetaXFunction}),
and applying the free-evolution operator produces in the case $\kbar=\pi$:\begin{widetext}
\begin{eqnarray}
\psi_{\beta}(X,t^{-}) & = & \frac{1}{\sqrt{2\pi}}\sum_{n}\widetilde{\psi}_{\beta}(n,t-1)\exp
\left(-i\frac{\pi}{2}\left(n+\beta\right)^{2}\right)e^{i(n+\beta)X}\nonumber \\
 & = & \frac{1}{\sqrt{2\pi}}\exp\left(-\frac{i\pi\beta^{2}}{2}\right)\sum_{n}
\widetilde{\psi}_{\beta}(n,t-1)\exp\left(-\frac{i\pi n^{2}}{2}\right)\exp
\left(-i\pi n\beta\right)\exp\left(i(n+\beta)X\right).\label{eq:pi-BW1}\end{eqnarray}
We show in the Appendix \ref{app:pi-Evolution} that the above expression
can be written as \begin{eqnarray}
\psi_{\beta}\left(X,t\right) & = & e^{i\pi\beta^{2}/2}\frac{e^{-i\kappa\cos X}}{\sqrt{2}}\left[e^{-i
\frac{\pi}{4}}\psi_{\beta}\left(X-\kbar\beta,t-1\right)+e^{i\frac{\pi}{4}}e^{i\beta\pi}
\psi_{\beta}\left(X-\kbar\beta-\pi,t-1\right)\right]\label{eq:pi-rec-psi}\end{eqnarray}
or\begin{eqnarray}
\psi_{\beta}(X+wt,t) & = & \frac{e^{-i\kappa\phi(X,t)}}{\sqrt{2}}\left[e^{-i\pi/4}
\psi_{\beta}\left(X+w(t-1),t-1\right)+e^{i\pi/4}e^{i\beta\pi}\psi_{\beta}\left(X+w(t-1)-
\pi,t-1\right)\right]\label{eq:pi-BWevolution}\end{eqnarray}
\end{widetext}where $w$ is the packet (stroboscopic) drift velocity
defined by \begin{equation}
w=\kbar\beta\label{eq:w-HQR}\end{equation}
and we introduced the {}``local'' phase \citep{note:firstfactor}
\begin{equation}
\phi(x,t)=\kappa\cos(X+wt)\label{eq:LocalPhase}\end{equation}
Eq.~(\ref{eq:pi-BWevolution}) shows that at any time $t$, $\psi_{\beta}(X+wt,t)$
is the superposition of two subpackets having the same shape as the
initial BW, centered at $X=0$ and $X=\pi$. Each of these two subpackets
is multiplied by a phase factor which is the sum of the accumulated
phases, producing a complex interference pattern. The principle of
the following calculation is to keep track of the coefficient of each
subpacket, as we know that the \emph{shape} of the subpackets is fixed.
We then write:
\begin{equation}
\psi_{\beta}(X+wt,t)=c_{1}(X,t)\psi_{\beta}(X,0)+c_{2}(X,t)\psi_{\beta}(X-\pi,0),
\label{eq:pi-decompo-psi}
\end{equation}
where $c_{1}(X,t)$ and $c_{2}(X,t)$ are $2\pi$-periodic complex
amplitudes \citep{note:periodicity}. The above expression corresponds
to a coupled two-level model where the {}``particle'', initially
in {}``level 1'' (i.e $c_{1}=1$ at $t=0)$ is progressively {}``transferred''
to level 2'' ($c_{2}\neq0)$ and then back again to {}``level 1'',
performing a kind of Rabi oscillation. 

Let us define the state vector \[
\mathbf{c}_{t}=\left(\begin{array}{c}
c_{1}(X,t)\\
c_{2}(X-\pi,t)\end{array}\right)\]
It is shown in App. \ref{appsub:"Mode-expansion"} that these amplitudes
obey a matrix recurrence relation : \[
\mathbf{c}_{t}=M_{t}\mathbf{c}_{t-1}\]
where $M_{t}$ is a matrix depending on time and space having the
form\[
M_{t}=e^{-i\pi/4}\widetilde{M}_{t}\]
 where \begin{equation}
\widetilde{M}_{t}=\frac{1}{\sqrt{2}}\left(\begin{array}{cc}
e^{-i\phi} & ie^{-i\phi}e^{-i\beta\pi}\\
ie^{i\phi}e^{i\beta\pi} & e^{i\phi}\end{array}\right),\label{eq:pi-Mt}\end{equation}
with $\phi(X+wt)$ given by Eq.~(\ref{eq:LocalPhase}). The matrix
$\widetilde{M}_{t}$ can be recast as\[
\widetilde{M}_{t}=\frac{1}{\sqrt{2}}\left(\begin{array}{cc}
e^{-i\phi} & 0\\
0 & e^{i\phi}\end{array}\right)\left(\begin{array}{cc}
1 & ie^{-i\beta\pi}\\
ie^{i\beta\pi} & 1\end{array}\right).\]
The rightmost matrix in the above product stands for the free propagation
that induces a coupling between the two subpackets; it is thus responsible
of the interference effects. The leftmost one represents the effect
of the kick and is obviously diagonal in $x$-representation. 

Analytical results can be obtained in the case $\beta=0$. As in this
case $w=0$, the matrix $\widetilde{M}_{t}$ is time-independent.
It is then easy to write $\mathbf{c}_{t}$ as a function of the initial
condition\[
\mathbf{c}_{0}=\left(\begin{array}{c}
1\\
0\end{array}\right):\]
 \begin{eqnarray*}
\mathbf{c}_{t} & =\left[M_{t}\right]^{t}\mathbf{c}_{0}=e^{-it\pi/4}
\left[\widetilde{M}_{t}\right]^{t} & \mathbf{c}_{0}\end{eqnarray*}
The eigenvalues of $\widetilde{M}_{t}$ are:\[
\lambda=\exp(\pm i\Theta)\]
where the phase $\Theta$ depends on $X$ and is given by\begin{eqnarray*}
\cos\Theta & = & \frac{\cos\phi}{\sqrt{2}}\\
 & = & \frac{\cos(\kappa\cos X)}{\sqrt{2}}\end{eqnarray*}
 (note that $\pi/4\leq\Theta\leq3\pi/4)$. If $\mathbf{P}$ is the
diagonalizing matrix, one can write (see Sec. \ref{appsub:beta=00003D0}
in App. \ref{app:pi-Evolution}) \begin{equation}
\mathbf{c}_{t}=e^{-it\frac{\pi}{4}}\mathbf{P}\left(\begin{array}{cc}
e^{it\Theta} & 0\\
0 & e^{-it\Theta}\end{array}\right)\mathbf{P}^{-1}\mathbf{c}_{0}.\label{eq:pi-produit}\end{equation}
After a straightforward calculation, the amplitudes are found to be
\begin{equation}
c_{1}(X,t)=e^{-it\frac{\pi}{4}}\left[\cos(t\Theta)-\frac{i\sqrt{2}}{2}
\frac{\sin\phi}{\sin\Theta}\sin(t\Theta)\right]\label{eq:pi-c1}\end{equation}
and\begin{equation}
c_{2}(X-\pi,t)=\frac{i\sqrt{2}}{2}\frac{e^{i\phi}}{\sin\Theta}e^{-it
\frac{\pi}{4}}\sin(t\Theta).\label{eq:pi-c2}\end{equation}
From the above result we can calculate the average momentum, but the
algebra involved is cumbersome (see App. \ref{app:pi-Averages}):\[
\left\langle P\right\rangle _{\beta}(t)=\left\langle P\right\rangle _{\beta}(t-1)+
K\int_{-\pi}^{\pi}dX\sin X\left|\psi_{\beta}(X,t)\right|^{2}.\]

In the general case, it is difficult to analyze
the behavior of the amplitudes $c_{1}(X,t)$ and $c_{2}(X,t)$ since
they depend implicitly on $X$ in a quite complicated way. Analytic
results can however be obtained in the limit which the initial BW
is formed of spatially narrow wavepackets.Let
us assume that $\psi_{\beta}(X,t=0)$ is well localized at some position
$X_{0}$ in the interval $\left[-\pi,\pi\right]$ (i.e its width $\Delta X\ll2\pi$)
and express the average momentum in terms of the coefficients $c_{1}$
and $c_{2}$\begin{widetext}
\begin{eqnarray}
\left\langle P\right\rangle _{\beta}(t) & = & \left\langle P\right
\rangle _{\beta}(t-1)+K\int_{-\pi}^{\pi}dx\sin(X+wt)
\left(\left|c_{1}(X,t)\right|^{2}-\left|c_{2}(X-\pi,t)
\right|^{2}\right)\left|\psi_{\beta}(X,0)\right|^{2}
\label{eq:P-ave-Beta-c1-c2}
\end{eqnarray}
\end{widetext}This expression shows that if the two subpackets have
the same weight, the momentum shift per kick is zero: the subpackets
are localized around positions $X_{0}$ and $X_{0}+\pi$, and subjected
to opposite forces $+K\sin(X_{0})/2$ and $-K\sin(X_{0})/2$, an effect
that is characteristic of the high-order resonances and obviously
does not exist for simple resonances. This expression is valid if
the two subpackets are well separated so that they do not interfere
significantly.

For a wavepacket that is strongly localized at $X_{0}$,
the amplitudes in the decomposition Eq.~(\ref{eq:P-ave-Beta-c1-c2})
can be evaluated at $X_{0}$ and depend only on time, while the phases
$\phi$ and $\Theta$ take constant values ($\phi=\kappa\cos X_{0})$.
Hence, the average momentum evolution is\begin{widetext}
\begin{equation}
\left\langle P\right\rangle _{\beta}(t)=\left\langle P\right\rangle _{\beta}(t-1)
+K\sin\left(X_{0}+wt\right)\left(1-2\left|c_{2}(X_{0}-\pi,t)\right|^{2}\right),
\label{eq:pi-rec-p}
\end{equation}
which, in our simple case $\beta=0$, takes the explicit form\[
\left\langle P\right\rangle _{\beta=0}(t)=\left\langle P\right\rangle _{\beta=0}(t-1)
+K\sin X_{0}\left(1-\frac{\sin^{2}(t\Theta)}{\sin^{2}\Theta}\right).\]
Note that the expression inside parenthesis in the above expression
is characteristic of the diffraction on a grating, which puts into
evidence the {}``wavelike'' nature of the dynamics. Iterating down
to $t=0$ produces \[
\left\langle P\right\rangle _{\beta=0}(t)=\left\langle P\right\rangle _{\beta=0}(t=0)+
K\sin X_{0}\left[\left(\frac{\sin^{2}\phi}{1+\sin^{2}\phi}\right)t+
\frac{1}{1+\sin^{2}\phi}\left(\frac{\sin\left[(2t+1)\Theta\right]}{2\sin\Theta}-
\frac{1}{2}\right)\right].\]
\end{widetext}This expression shows that average momentum evolution
is, for arbitrary $\phi$ and $\Theta$, a \emph{mix} of two qualitatively
different behaviors: \emph{ballisticity}, corresponding to the first
term in the brackets, and \emph{oscillation}, described by the second
term. This contrasts with SQRs, where the dynamics is \emph{either}
ballistic or oscillatory.

\begin{figure}
\begin{centering}
\includegraphics[bb=50bp 30bp 410bp 302bp,clip,width=8cm]{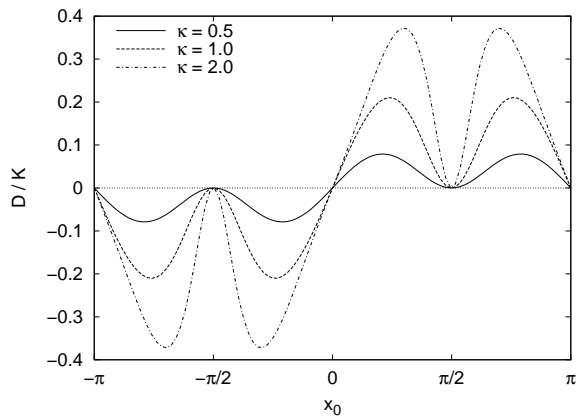} 
\par\end{centering}

\caption{\label{fig:Dx0}Momentum slope $D$ {[}Eq.~(\ref{eq:pi-diffusion})
] as a function of $X_{0}$ for 3 values of $\kappa$: 0.5 (full line),
1.0 (dots) and 2.0 (dashed).}
\end{figure}

For long times and $\phi\neq0$, the ballistic term dominates and
one gets \[
\left\langle P\right\rangle _{\beta=0}(t)=\left\langle P\right\rangle _{\beta=0}(t=0)+Dt\]
with \begin{equation}
D=K\sin X_{0}\left(\frac{\sin^{2}\phi}{1+\sin^{2}\phi}\right).\label{eq:pi-diffusion}\end{equation}
being the rate of change of the mean momentum. The behavior of $D$
as a function of $X_{0}$ is displayed in Fig.~\ref{fig:Dx0} for
different values of $\kappa$. It is interesting to compare Eq.~(\ref{eq:pi-diffusion})
with its counterpart for SQRs, which is given by Eq.~(\ref{eq:2pil-D}):
note in particular that in the present case $D=0$ if $X_{0}=\pi/2$,
whereas for SQR, the maximum of $D$ occurs when the force is maximum
(i.e at $X_{0}=\pi/2)$.

\begin{figure}
\begin{centering}
\includegraphics[bb=50bp 30bp 410bp 302bp,clip,width=8cm]{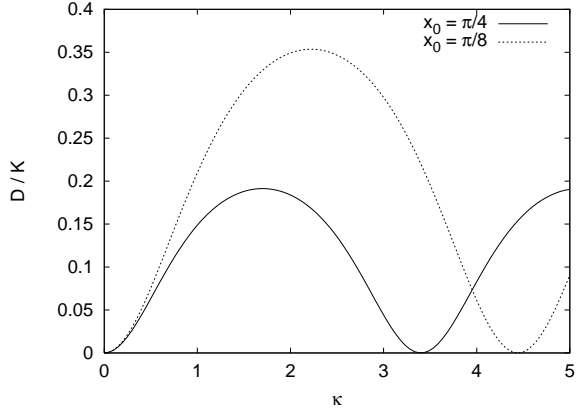}
\caption{\label{fig:Dkappa} Momentum slope $D$ {[}Eq.~(\ref{eq:pi-diffusion})]
as a function of $\kappa$ for two initial positions $X_{0}$=$\pi/4$
(dots) and $X_{0}=\pi/8$ (full line). One sees that increasing $\kappa$
(that is, the kick intensity) does not necessarily increase the momentum
slope.}
\par\end{centering}
\end{figure}

The coefficient $D$ depends periodically on the kick intensity $\kappa$
\emph{via} $\phi=\kappa\cos X_{0}$, as shown in Fig.~\ref{fig:Dkappa}.
In contrast to the simple resonance case, increasing the kick force
does not necessarily increase the diffusion, an effect that persists
if the momentum is averaged over the quasimomentum distribution.

Analogous results describing the ballistic behavior of an initial
state which is an eigenvector of the momentum have been obtained via
a quite different approach \citep{Casati:QuatumResAnalytic:PRE00}.
Our localized-packet approach provides a clearer picture of the underlying
physics.

\begin{figure}
\begin{centering}
\includegraphics[clip,width=8cm]{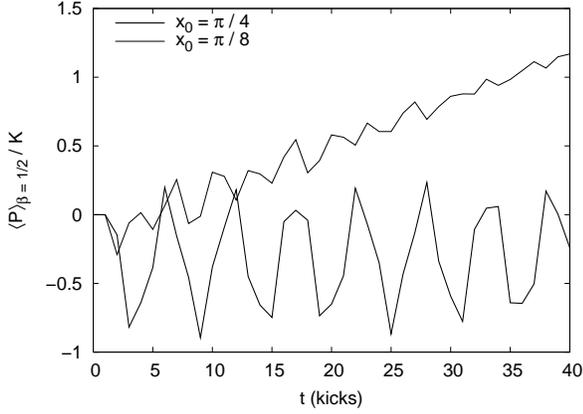}
\par\end{centering}
\caption{\label{fig:Paveragebeta0.5} Evolution of the average momentum for
$\beta=1/2$ obtained by numerical integration of the Schrödinger
equation. The initial initial wavefunction has the same Gaussian shape
as in Fig.~\ref{fig:FreeEvolution} and is centered at the positions
$X_{0}=\pi/4$ (dark line), which displays a dominant oscillatory
behavior, or $X_{0}=\pi/8$ (light line), displaying dominant ballisticity.}
\end{figure}

For $\beta\neq0$, the occurrence of ballisticity depends on the periodicity
of $\text{M}_{t}$ and on the initial conditions. More precisely,
ballisticity will emerge if the relation $M_{t+t_{r}}=M_{t}$ is fulfilled
for some (integer) recurrence time $t_{r}$. This happens for any
\emph{rational} value of quasimomentum. For example, if $\beta=1/2$,
$\text{M}_{t}$ {[}see Eq.~(\ref{eq:pi-Mt})] has a period of 4 kicks:
$\text{M}_{t+4}=\text{M}_{t}$. One can apply the above reasoning
to $\text{M}=\text{M}_{4}\text{M}_{3}\text{M}_{2}\text{M}_{1}$. To
illustrate the resulting behavior, Fig.~\ref{fig:Paveragebeta0.5}
shows the time-evolution of average momentum obtained by direct integration
of the Schrödinger equation. One observes essentially the same kind
of behavior.

For a general rational quasimomentum $\beta=p/q$, the ballistic diffusion
rate is roughly proportional to $1/q$. Irrational quasimomenta, that
may be consider as the limit $q\rightarrow+\infty$, do not produce
ballistic behavior.

\section{Averaging over quasimomentum}

In experiments performed with laser-cooled atoms
(not with \emph{ultracold}atoms
-- BECs) the initial momentum distribution is larger than the Brillouin
zone unless velocity selection is performed. In position space, this
corresponds to an initial wavepacket which is localized on a single
potential well, around a position $X_{0}$. In
this simple case, and for $\kbar=2\pi\ell$, we are able to give
an analytical expression for any value of $t$, of the momentum and
kinetic energy averaged on quasimomentum $\beta$.

Starting from Eq.~(\ref{eq:moyenne-P-psi}):
\begin{widetext}\[
\langle P\rangle(t)=\langle P\rangle(t=0)+K\sum_{n=1}^{t}\int_{-1/2}^{1/2}d\beta
\int_{-\pi}^{\pi}dX\sin(X+nv)\left|\psi_{\beta}(X,0)\right|^{2}.\]

Assuming that $\psi_{\beta}(X,0)$ is $\beta-$independent ($\psi_{\beta}(X,0)=\varphi(X)$and
$\int_{-\pi}^{\pi}dx\left|\varphi(X)\right|^{2}=1$ ), and recalling
that $v=2\pi\ell(\beta+1/2)$, we easily see that averaging on $\beta$
lead to:\[
\left\langle P\right\rangle (t)=\left\langle P\right\rangle (t=0).\]
A similar reasoning can be applied for kinetic energy. From Eqs. (\ref{eq:moyenne-E-psi})
and (\ref{eq:2pil-rec-ener}), we have:
\begin{eqnarray*}
\langle E\rangle(t) & = & \langle E\rangle(t=0)+\frac{K^{2}}{2}\int_{-1/2}^{1/2}d\beta
\int_{-\pi}^{\pi}dX\left(\sum_{n=1}^{t}\sin(X+nv)\right)^{2}\left|\varphi(X)\right|^{2}
+K\sum_{n=1}^{t}\int_{-1/2}^{1/2}d\beta\int_{-\pi}^{\pi}dx\sin(X+nv)J(X,t=0)
\end{eqnarray*}
When we integrate over $\beta$, the term of the last term cancels
out, and the only contributions come from \[
\int_{-1/2}^{1/2}d\beta\left(\sum_{n=1}^{t}\sin(X+nv)\right)^{2}=\int_{-1/2}^{1/2}
d\beta\left(\sum_{n=1}^{t}\sin^{2}(X+\pi\ell n+2\pi\ell n\beta)\right)=\frac{t}{2}\]
\end{widetext}We finally get the diffusive behavior
\begin{equation}
\langle E\rangle(t)=\langle E\rangle(t=0)+\frac{K^{2}t}{4}.\label{eq:2pil-ener-diffus}
\end{equation}

This result evokes the classical kicked rotor in
chaotic regime, whose kinetic energy grows linearly with time with
the same rate. This shows that the ballistic behavior, which corresponds
to a ``null-measure ensemble'' of rational quasimomenta, is very
hard to detect by measuring quantities averaged over quasimomentum.

Experimentally, the optimum situation for observing
QRs is to perform a quasimomentum selection, either by using stimulated
Raman transitions \citep{Steinberg:HighOrderQRes:PRL07,Chu:RamanCooling:PRL92,AP:Polarization:OC07}
or by using a Bose-Einstein condensate \citep{Phillips:HighOrderQuantResBEC:PRL06,
DArcy:HighOrderQRes:PRL03,Leonhardt:KREarlyTimeDiff:PRL04}.
However, it is possible to detect QRs with atoms issued of a magneto-optical
trap if one can measure the full momentum distribution with enough
precision to see the ballistic parts of the wavefunction separating
out of the diffusive part for long enough times, as experimentally
evidenced by d'Arcy \emph{et al.} \citep{DArcy:AccModes:PRL99}.

\section{Conclusion\label{sec:Conclusion}}

In the present work, we presented a description of quantum resonances
of the kicked rotor in position space, both for simple and for high-order
quantum resonances. We have shown that the dynamics can be understood
by considering that the spatial wavepacket comes back to its in ital
form after a finite number of kicks, according to the (rational) value
of the quasimomentum. For a localized wavepacket (or an finite ensemble
of localized wavepackets, in the case of HQRs), one can interpret
the dynamical behavior in terms of the action of a finite number of
successive kicks. This picture, inspired of the atom-optics analog
of the Talbot effect, proves very useful both as it providing a intuitive
understanding of the underlying physics ans as it leads to analytical
developments for experimentally-relevant quantities.

\appendix

\section{Average momentum\label{app:mean}}

The average momentum reads:

\[
\left\langle P\right\rangle =\kbar\sum_{n}\int_{-1/2}^{1/2}d\beta(n+\beta)\left\vert 
\widetilde{\psi}_{\beta}\left(n\right)\right\vert ^{2}\]
and can be related to $\left\langle P\right\rangle _{\beta}$ in the
following way. One has from the definition of section \ref{sec:Physical-origin}:

\[
\widetilde{\psi}_{\beta}\left(n\right)=\frac{1}{\sqrt{2\pi}}\int_{-\pi}^{\pi}
\psi_{\beta}\left(X\right)e^{-i(\beta+n)X}dx\]
therefore:\begin{widetext}\[
(n+\beta)\widetilde{\psi}_{\beta}\left(n\right)=\frac{i}{\sqrt{2\pi}}
\int_{-\pi}^{\pi}dX\psi_{\beta}\left(X\right)\frac{\partial}{\partial X}e^{-i(\beta+n)X}
=\frac{i}{\sqrt{2\pi}}\left[\psi_{\beta}\left(X\right)
e^{-i(\beta+n)X}\right]_{-\pi}^{\pi}-\frac{i}{\sqrt{2\pi}}\int_{-\pi}^{\pi}dX
e^{-i(\beta+n)X}\frac{\partial}{\partial X}\psi_{\beta}\left(X\right)\]
The first term on the RHS vanishes ($u_{\beta}\left(X\right)=\psi_{\beta}
\left(X\right)e^{-i\beta X}$
is $2\pi-$periodic). One then has:
\begin{eqnarray*}
\left\langle P\right\rangle  & = & -\frac{i\kbar}{\sqrt{2\pi}}\sum_{n}
\int d\beta\widetilde{\psi}_{\beta}^{\ast}\left(n\right)
\int_{-\pi}^{\pi}dXe^{-i(\beta+n)X}\frac{\partial}{\partial X}\psi_{\beta}\left(X\right)\\
 & = & -\frac{i\kbar}{2\pi}\int d\beta\int_{-\pi}^{\pi}
\int_{-\pi}^{\pi}dXdX^{\prime}e^{-i\beta(X-X^{\prime})}
\psi_{\beta}\left(X^{\prime}\right)\frac{\partial}{\partial X}
\psi_{\beta}\left(X\right)\sum_{n}e^{-in(X-X^{\prime})}
\end{eqnarray*}
Finally, using $\sum_{n}e^{-in(X-X^{\prime})}=\sum_{k}\delta(X-X^{\prime}-2k\pi)$\[
\left\langle P\right\rangle =-i\kbar\int d\beta\int_{-\pi}^{\pi}dX\psi_{\beta}^{\ast}
\left(X\right)\frac{\partial}{\partial X}\psi_{\beta}\left(X\right)=
\int_{-1/2}^{1/2}d\beta\left\langle P\right\rangle _{\beta}\]
where \[
\left\langle P\right\rangle _{\beta}=\int_{-\pi}^{\pi}dX\psi_{\beta}^{\ast}\left(X\right)
\left[P\psi_{\beta}\left(X\right)\right]\]
For the kinetic energy, the same developments lead to:\[
\left\langle P^{2}\right\rangle =\sum_{n}\int d\beta(n+\beta)^{2}\left\vert
 \widetilde{\psi}_{\beta}\left(n\right)\right\vert ^{2}=
\int d\beta\left\langle P^{2}\right\rangle _{\beta}\]
where \[
\left\langle P^{2}\right\rangle _{\beta}=\int_{-\pi}^{\pi}dX
\left[P\psi_{\beta}\left(X\right)\right]^{\ast}\left[P\psi_{\beta}
\left(X\right)\right]\]
(note that the same rules can be obtained for $\left\langle P^{k}\right\rangle =
\int d\beta\left\langle P^{k}\right\rangle _{\beta}$).

\section{Bloch wave evolution for the $\kbar=\pi$ resonance\label{app:pi-Evolution}}

\subsection{Bloch wave evolution}

The free-propagation factor $\exp(-i\frac{\pi}{2}n^{2})$ in Eq.~(\ref{eq:pi-BW1})
has two different values according to the parity of $n$:\begin{eqnarray*}
\exp\left(-i\frac{\pi}{2}n^{2}\right) & = & 1\;\text{ (\emph{n} even)}\\
 & = & -i\text{\; (\emph{n} odd)}.\end{eqnarray*}
Replacing\[
\widetilde{\psi}_{\beta}(n,t)=\frac{1}{\sqrt{2\pi}}\int_{-\pi}^{\pi}dX
e^{-inX}u_{\beta}(X,t)=\frac{1}{\sqrt{2\pi}}
\int_{-\pi}^{\pi}dXe^{-i(n+\beta)X}\psi_{\beta}(X,t)\]
 in Eq.~(\ref{eq:pi-BW1}), one finds\[
\psi_{\beta}(X,t^{-})=\frac{\exp\left(-i\pi\beta^{2}/2\right)}{2\pi}\sum_{n}
\int_{-\pi}^{\pi}dX^\prime e^{-i(n+\beta)(X-X^\prime)}\psi_{\beta}(X^\prime,t-1)e^{-i\pi n^{2}/2}e^{-i\pi n\beta}\]
 one can separate even and odd terms ; this leads to:\[
\psi_{\beta}(X,t^{-})=\frac{\exp\left(-i\pi\beta^{2}/2\right)}{2\pi}
\int_{-\pi}^{\pi}dX^\prime\psi_{\beta}(X^\prime,t-1)e^{-i\beta(X-X^\prime)}
\left[\sum_{p}e^{-i2\pi p\beta}e^{-i2p(X-X^\prime)}\left(1-ie^{-i\pi\beta}
e^{-i(X-X^\prime)}\right)\right]\]
Using the relation $\sum_{p}e^{-i2\pi p\beta}e^{-i2p(X-X^\prime)}=(1/2)
\sum_{n}\delta(X-X^\prime-\pi\beta+n\pi)$
and integrating with respect to $X^\prime$, gives Eq.~(\ref{eq:pi-BWevolution})
(note that only $n=0,1$ contributes for $\beta>0$ and $n=0,-1$
for $\beta$<0).

\subsection{{}``Two-level'' system \label{appsub:"Mode-expansion"}}

The coupled equation for the amplitudes $c_{1,2}(X,t)$ are obtained
in the following way. Insertion of Eq.~(\ref{eq:pi-decompo-psi})
at time $(t-1)$

\begin{equation}
\psi_{\beta}(X+w(t-1),t-1)=c_{1}(X,t-1)\psi_{\beta}(X,0)+c_{2}(X,t-1)\psi_{\beta}(X-\pi,0),
\label{eq:app_mode_expansion}
\end{equation}
in Eq.~(\ref{eq:pi-BWevolution}) gives 
\begin{eqnarray*}
\psi_{\beta}(X+wt,t) & = & \frac{e^{-i\phi(X,t)}}{\sqrt{2}}\left(e^{-i\pi/4}
\left[c_{1}(X,t-1)\psi_{\beta}(X,0)+c_{2}(X,t-1)\psi_{\beta}(X-\pi,0)\right]+\right.\\
 &  & \left.e^{i\pi/4}e^{i\beta\pi}\left[c_{1}(X-\pi,t-1)\psi_{\beta}(X-\pi,0)+
c_{2}(X-\pi,t-1)\psi_{\beta}(X-2\pi,0)\right]\right)
\end{eqnarray*}
and can be put in a simpler form {[}using $\psi_{\beta}(X-2\pi,0)
=e^{-i2\pi\beta}\psi_{\beta}(X,0)$]:
\begin{eqnarray*}
\psi_{\beta}(X+wt,t) & = & \frac{e^{-i\phi(X,t)}}{\sqrt{2}}\left(\left[e^{-i\pi/4}c_{1}(X,t-1)
+e^{i\pi/4}e^{-i\beta\pi}c_{2}(X-\pi,t-1)\right]\psi_{\beta}(X,0)+\right.\\
 &  & \left.\left[e^{i\pi/4}e^{i\beta\pi}c_{1}(X-\pi,t-1)+e^{-i\pi/4}c_{2}(X,t-1)
\right]\psi_{\beta}(X-\pi,0)\right)
\end{eqnarray*}
Comparing to Eq.~(\ref{eq:pi-decompo-psi}) one obtains:
\begin{eqnarray*}
c_{1}(X,t) & = & \frac{e^{-i\phi(X,t)}}{\sqrt{2}}\left[e^{-i\pi/4}c_{1}(X,t-1)
+e^{i\pi/4}e^{-i\beta\pi}c_{2}(X-\pi,t-1)\right]\\
c_{2}(X,t) & = & \frac{e^{-i\phi(X,t)}}{\sqrt{2}}\left[e^{i\pi/4}
e^{i\beta\pi}c_{1}(X-\pi,t-1)+e^{-i\pi/4}c_{2}(X,t-1)\right].
\end{eqnarray*}
This last expression can be put into the form \[
c_{2}(X-\pi,t)=\frac{e^{i\phi(X,t)}}{\sqrt{2}}
\left[e^{i\pi/4}e^{i\beta\pi}c_{1}(X-2\pi,t-1)+e^{-i\pi/4}c_{2}(X-\pi,t-1)\right].\]
One can show that the amplitudes are periodic functions, i.e $c_{1,2}(X,t)$$=c_{1,2}(X+2\pi,t)$
by combining Eq.~(\ref{eq:pi-decompo-psi}) combined with the equality
$\psi_{\beta}(X-2\pi,t)$$=e^{-i2\pi\beta}\psi_{\beta}(X,t)$. This
property leads to the matrix expression\[
\left(\begin{array}{c}
c_{1}(X,t)\\
c_{2}(X+\pi,t)\end{array}\right)=\frac{1}{\sqrt{2}}\left(\begin{array}{cc}
e^{-i\phi}e^{-i\pi/4} & e^{-i\phi}e^{i\pi/4}e^{-i\beta\pi}\\
e^{i\phi}e^{i\pi/4}e^{i\beta\pi} & e^{i\phi}e^{-i\pi/4}\end{array}\right)\left(\begin{array}{c}
c_{1}(X,t-1)\\
c_{2}(X+\pi,t-1)\end{array}\right)\]

\subsection{The case $\beta=0$\label{appsub:beta=00003D0}}

The explicit expression for the amplitudes $c_{1,2}(X,t)$ is obtained
for $\beta=0$ as follows. The diagonalization matrix $\mathbf{P}$
is formed with the eigenvectors of $\widetilde{M}_{t}$. For $\lambda=e^{\pm i\Theta}$
they are given by$\begin{array}{cc}
(-ie^{-i\phi}, & e^{-i\phi}-\sqrt{2}e^{\pm i\Theta})^{T}\end{array}$ . 
$\mathbf{P}$ is then obtained as \[
\mathbf{P}=\left[\begin{array}{cc}
-ie^{-i\phi} & -ie^{-i\phi}\\
e^{-i\phi}-\sqrt{2}e^{i\Theta} & e^{-i\phi}-\sqrt{2}e^{-i\Theta}\end{array}\right].\]
 In order to obtain the amplitudes $c_{1}$ and $c_{2}$ at time $t$,
one performs explicitly the development corresponding to Eq.~(\ref{eq:pi-produit}).
The algebra is simple, although rather long, and the final result
is Eq.~(\ref{eq:pi-c1}).

\section{Calculus of average values for the $\kbar=\pi$ resonance\label{app:pi-Averages}}

Starting from Eq.~(\ref{eq:pi-rec-psi}), recursion relation for $\psi_{\beta}\left(X,t\right)$,
we easily obtain a recursion relation for its derivative
\begin{eqnarray}
\frac{\partial}{\partial X}\psi_{\beta}\left(X,t\right) & = & 
i\kappa\sin Xe^{-i\kappa\cos X}\psi_{\beta}\left(X,t\right)\notag\\
 & + & \frac{e^{-i\kappa\cos X}}{\sqrt{2}}\left(e^{-i\frac{\pi}{4}}
\frac{\partial}{\partial X}\psi_{\beta}\left(X-w,t-1\right)+
e^{i\frac{\pi}{4}}e^{i\beta\pi}\frac{\partial}{\partial X}\psi_{\beta}
\left(X-w-\pi,t-1\right)\right).\label{eq:pi-rec-dpsi-beta}
\end{eqnarray}
Using these expressions into for calculating the average momentum
produces two terms; let us call them $p_{1}$ and $p_{2}$. The first
one is given by:
\begin{eqnarray*}
p_{1} & = & K\int_{-\pi}^{\pi}dX\sin X\left|\psi_{\beta}(X,t)\right|^{2}
\end{eqnarray*}
 For the second term $p_{2}$, on obtains
\begin{eqnarray*}
p_{2} & = & -\frac{i\kbar}{2}\int_{-\pi}^{\pi}dX
\left(e^{i\frac{\pi}{4}}\psi_{\beta}^{*}\left(X-w,t-1\right)+
e^{-i\frac{\pi}{4}}e^{-i\beta\pi}\psi_{\beta}^{*}
\left(X-x-\pi,t-1\right)\right)\\
 &  & \times\left(e^{-i\frac{\pi}{4}}\frac{\partial}{\partial X}
\psi_{\beta}\left(X-w,t-1\right)+e^{i\frac{\pi}{4}}e^{i\beta\pi}
\frac{\partial}{\partial X}\psi_{\beta}\left(X-w-\pi,t-1\right)\right)\\
 & = & -\frac{i\kbar}{2}\int_{-\pi}^{\pi}dX\left(\psi_{\beta}^{*}
\left(X,t-1\right)\frac{\partial}{\partial X}\psi_{\beta}
\left(X,t-1\right)+\psi_{\beta}^{*}\left(X-\pi,t-1\right)
\frac{\partial}{\partial X}\psi_{\beta}\left(X-\pi,t-1\right)\right)\\
 & + & \frac{\kbar}{2}\int_{-\pi}^{\pi}dX\left(e^{i\beta\pi}
\psi_{\beta}^{*}\left(X,t-1\right)\frac{\partial}{\partial X}
\psi_{\beta}\left(X-\pi,t-1\right)-e^{-i\beta\pi}\psi_{\beta}^{*}
\left(X-\pi,t-1\right)\frac{\partial}{\partial X}\psi_{\beta}
\left(X,t-1\right)\right)
\end{eqnarray*}
{[}in the last equality we replaced $(X-w)$by $X$, as the integration
is over one period, the integration limits can be kept the same].
One recognizes $\left\langle P\right\rangle _{\beta}(t-1)$ in the
first term while the second term cancels out. The recursion relation
for the momentum thus reads
\begin{equation}
\left\langle P\right\rangle _{\beta}(t)=\left\langle P\right
\rangle _{\beta}(t-1)+K\int_{-\pi}^{\pi}dX\sin X\left|\psi_{\beta}(X,t)
\right|^{2}.\label{eq:pi-rec-P-1}
\end{equation}
We now use the decomposition of $\psi_{\beta}(X,t)$ proposed in Eq.
(\ref{eq:pi-decompo-psi}). Eq.~(\ref{eq:pi-rec-P-1}) then transforms
into:
\begin{eqnarray*}
\left\langle P\right\rangle _{\beta}(t) & = & \left\langle P\right
\rangle _{\beta}(t-1)+iK\cos(\beta\pi)\int_{-\pi}^{\pi}dX\sin(X+wt)
\left(c_{1}^{*}(X,t)\psi_{\beta}^{*}(X,0)+c_{2}^{*}(X,t)
\psi_{\beta}^{*}(X-\pi,0)\right)\\
 &  & \times\left(c_{1}(X,t)\psi_{\beta}(X,0)+c_{2}(X,t)
\psi_{\beta}(X-\pi,0)\right).
\end{eqnarray*}
If the initial wavefunction has a narrow distribution centered around
position $X_{0}$, all terms involving overlaps of functions $\psi_{\beta}(X,0)$
and $\psi_{\beta}(X+\pi)$ tend to zero \citep{note:narrow}. This
expression therefore simplifies to
\begin{eqnarray}
\left\langle P\right\rangle _{\beta}(t) & = & \left\langle P\right
\rangle _{\beta}(t-1)+K\int_{-\pi}^{\pi}dX\sin(X+wt)\left(\left|c_{1}(X,t)
\right|^{2}-\left|c_{2}(X-\pi,t)\right|^{2}\right)\left|\psi_{\beta}(X,0)
\right|^{2}\label{eq:Pave-beta}
\end{eqnarray}
Finally use the fact that $\psi_{\beta}(X,0)$ is much narrower then
any other factor, and assuming $\int_{-\pi}^{\pi}dX\left|\psi_{\beta}(X,0)\right|^{2}=1$
\citep{note:norm} gives\[
\left\langle P\right\rangle _{\beta}(t)=\left\langle P\right\rangle _{\beta}(t-1)
+K\sin(X_{0}+wt)\left(1-2\left|c_{2}(X_{0}-\pi,t)\right|^{2}\right)\]
where we used the normalization $\left|c_{1}(X,t)\right|^{2}
+\left|c_{2}(X-\pi,t)\right|^{2}=1$.\end{widetext}


\begin{thebibliography}{49}
\bibitem[{not({\natexlab{a}})}]{note:localized}
\bibinfo{note}{{One must however be careful about the meaning of a wavepacket
  when going from one representation to the other. To give an example that will
  play a role in the following, consider a wave packet that is ``localized''
  around a given position -- say $\theta_0$ -- in the folded representation. In
  the unfolded representation such a packet corresponds in fact to a ``comb''
  (of spatial step $\lambda_L/2$) of packets periodically distributed in $X$,
  with period $2\pi$.}}
\bibitem[{not({\natexlab{b}})}]{note:conventions}
\bibinfo{note}{{Throughout this work, we shall use the convention that $t$
  indicates the time immediately after the $t^{th}$ kick. When necessary, we
  shall use the notation $t^{+}$ to stress the fact that we are considering the
  instant after the $t^{th}$ kick has been applied, and $t^{-}$ to indicate the
  instant just before the $t^{th}$ kick is applied.}}
\bibitem[{not({\natexlab{c}})}]{note:combs}
\bibinfo{note}{{Indeed, a wavepacket can be spatially localized in the first
  Brillouin zone and can also have a well-defined quasimomentum. Both in the
  (unfolded) real space and in the momentum space the corresponding
  wavefunction is a periodic distribution of packets (an infinite comb) and the
  Heisenberg principle is fulfilled.}}
\bibitem[{not({\natexlab{d}})}]{note:drift}
\bibinfo{note}{{Note that the wavepacket in not simply drifting. For an
  arbitrary time between kicks it is completely delocalized (cf.
  Fig.~\protect\ref{fig:FreeEvolution}). It is only because we are studying the
  wavepacket at ``stroboscopic'' times $t$ in the vicinity of the kick,
  together with the ``refocalizing'' effect of the quantum resonance at these
  times that gives the impression of a drift. The true evolution for arbitrary
  times is much more complex.}}
\bibitem[{not({\natexlab{e}})}]{note:factor}
\bibinfo{note}{{The constant phase factor in
  Eq.~(\protect\ref{eq:2pil-rec-psi}) has been omitted in
  Eq.~(\protect\ref{eq:psi-beta-evolution}) for the sake of simplicity.}}
\bibitem[{not({\natexlab{f}})}]{note:firstfactor}
\bibinfo{note}{{The first constant phase factor in
  Eq.~(\protect\ref{eq:pi-rec-psi}) has been omitted for simplicity.}}
\bibitem[{not({\natexlab{h}})}]{note:narrow}
\bibinfo{note}{{Note that we cannot use the narrow wavepacket approximation
  directly in the calculation of the coefficients $c_{1,2}$ because in such
  case the derivative in Eq.~(\protect\ref{eq:pi-rec-dpsi-beta}) is
  meaningless. We can however use it in the expression of the average momentum
  thanks to the presence of the integral of $x$.}}
\end{thebibliography}

\end{document}